\documentclass{article}

\usepackage[utf8]{inputenc}
\usepackage{graphicx}
\usepackage{amsfonts}
\usepackage{subcaption}
\usepackage{url}
\usepackage{float}
\usepackage[normalem]{ulem}
\usepackage[labelfont=bf, font=scriptsize]{caption}
\usepackage{placeins}
\usepackage{authblk}
\usepackage[a4paper, total={6.3in, 10in}]{geometry}
\usepackage[colorlinks=true, citecolor=blue, filecolor=blue, linkcolor=black]{hyperref}
\usepackage{bbm}
\usepackage{color}
\usepackage{epstopdf}
\usepackage{comment}
\usepackage{multicol}
\usepackage{amsmath} 
\usepackage{color}
\usepackage{mathtools}
\usepackage[T1]{fontenc}    
\usepackage{url}            
\usepackage{booktabs}       
\usepackage{amsfonts}       
\usepackage{nicefrac}       
\usepackage{microtype}      
\usepackage{lipsum}
\usepackage{xcolor,colortbl}
\usepackage{amsmath}
\usepackage[numbers,sort&compress]{natbib}
\usepackage{lastpage,fancyhdr,graphicx}

\title{\textbf{Temporal origin of nestedness in interaction networks}}

\author{Phillip P. A. Staniczenko,$^{1,\ast}$ Debabrata Panja$^{2,3}$\\
  \normalsize{$^1$City University of New York (CUNY), Brooklyn College,}\\ \normalsize{2900 Bedford Avenue, Brooklyn, NY 11210, USA}\\
\normalsize{$^2$Department of Information and Computing Sciences, Utrecht University,}\\ \normalsize{Princetonplein 5, 3584 CC Utrecht, The Netherlands}\\
\normalsize{$^3$Centre for Complex Systems Studies, Utrecht University,}\\ \normalsize{Minnaertgebouw, Leuvenlaan 4, 3584 CE Utrecht, The Netherlands}\\
\normalsize{$^\ast$To whom correspondence should be addressed; E-mail:  pstaniczenko@brooklyn.cuny.edu}}
\date{}

\begin{document}
\maketitle

\begin{abstract}
\noindent Nestedness is a common property of communication, finance, trade, and ecological networks. In networks with high levels of nestedness, the link positions of low-degree nodes (those with few links) form nested subsets of the link positions of high-degree nodes (those with many links), leading to matrix representations with characteristic upper-triangular or staircase patterns. Recent theoretical work has connected nestedness to the functionality of complex systems and has suggested it is a structural by-product of the skewed degree distributions often seen in empirical data. However, mechanisms for generating nestedness remain poorly understood, limiting the connections that can be made between system processes and observed network structures. Here, we show that a simple probabilistic model based on phenology --- the timing of co-presences among interaction partners --- can produce nested structures and correctly predict around two-thirds of interactions in two fish market networks and around one-third of interactions in 22 plant-pollinator networks. Notably, the links most readily explained by frequent actor co-presences appear to form a backbone of nested interactions, with the remaining interactions attributable to opportunistic interactions or preferences for particular interaction partners that are not routinely available.

\vspace{5mm}
\noindent{\bf Significance statement:} Networks describe the relationships among actors in complex systems. In nested networks, actors involved in few interactions are connected to actors involved in many interactions, with those highly-connected actors also interacting with other highly-connected actors. This pattern is seen in a variety of empirical systems and influences the response to external perturbations, but little is known about the processes that give rise to nestedness. We show that phenology, the day-to-day timing of interaction partner availability, is a general mechanism that generates nested structures. We present a simple probabilistic model which accounts for actor overlap through time but assumes actors have no preference for specific interaction partners, thereby providing an instructive baseline for investigating higher-level selection processes in interaction networks.
\end{abstract}

\clearpage
\begin{figure*}
\begin{center}
\includegraphics[width=0.8\linewidth]{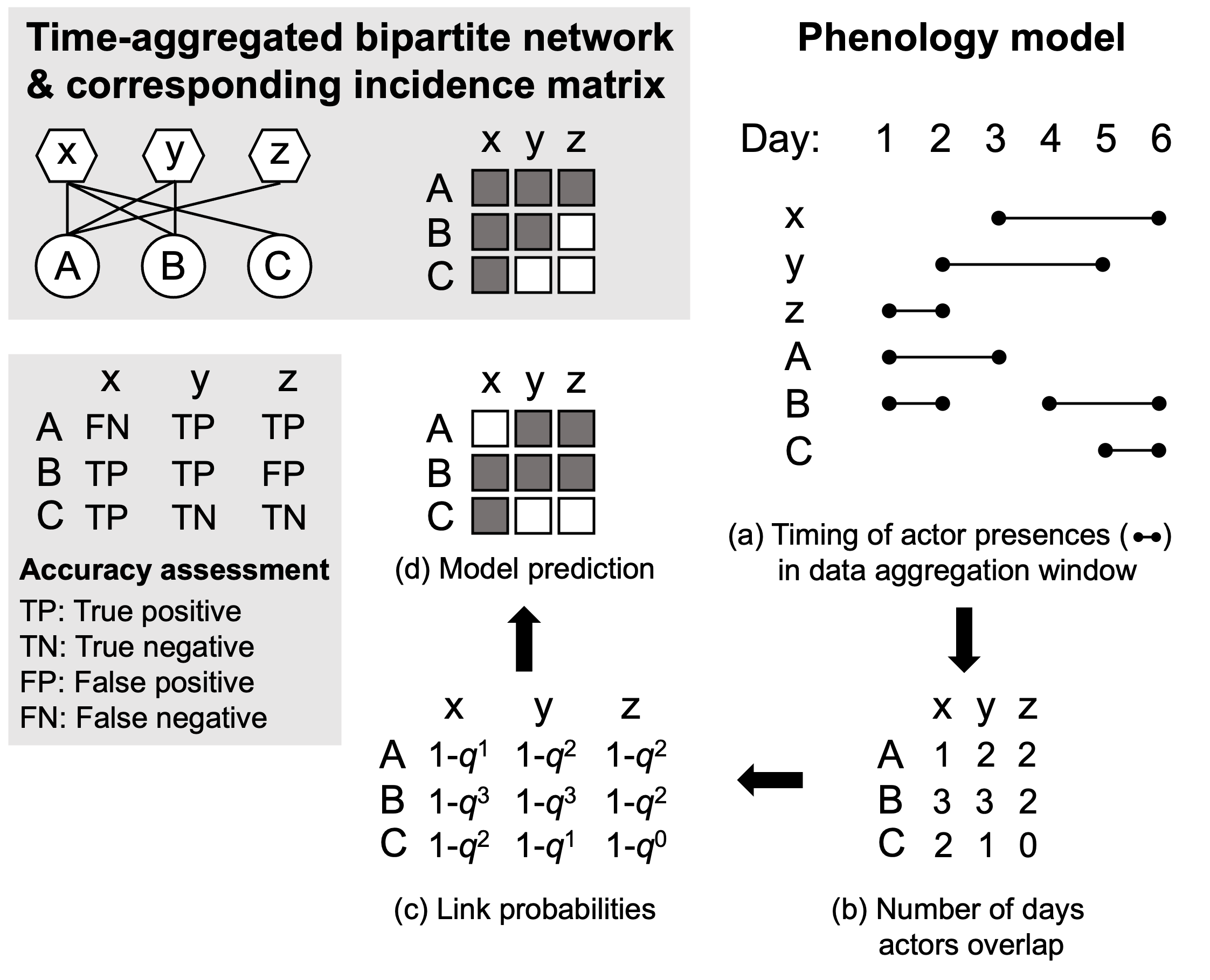}
\end{center}
\caption{Example of a perfectly nested bipartite network (upper box), outline of phenology model, and model assessment scheme (lower box). A link is included in an empirical network if an interaction is recorded during the data aggregation window between two actors of different types, node sets \{A,B,C\} and \{x,y,z\}, representing, e.g., different plant and pollinator species, respectively. In the corresponding incidence matrix, filled entries indicate links between nodes (rows and columns) and nestedness is characterized by an upper-triangular or staircase pattern, as shown. The phenology model starts with a description of when actors are present and active in the system during the data aggregation window, which in this case is 6 days (a). This temporal information dictates the number of co-presences between each pair of nodes (b), which is used to determine the probability of each link in model-generated networks (c). Specifically, the probability of an interaction given a co-presence is assumed to be $p$ and the probability of no interaction $q=1-p$, and so the probability of at least one interaction during the data aggregation window given $n$ co-presences is $1-q^n$. Link probabilities are scaled such that the expected number of links for an ensemble of model-generated networks equals the number in the empirical network ({\it Methods}). Each model realization returns an incidence matrix (d) and each link can assigned to one of four categories: true positive (TP) --- link present in both model and empirical networks; true negative (TN) --- link absent in both model and empirical networks; false positive (FP) --- link present in model network but absent in empirical network; false negative (FN) --- link absent in model network but present in empirical network.
\label{Figure1}}
\end{figure*}

\noindent The concept of nestedness first gained popularity in the field of island biogeography, where it was shown that the species found on distant islands formed progressive subsets of the species found on islands closer to the mainland~\cite{BiogeogNestedness}. Nestedness has since been studied in a wide variety of networks, most notably in bipartite networks in which links represent relationships between two different categories of actors~\cite{NestednessReview} (an illustration of the characteristic pattern of nestedness is shown in the upper-left panel in Fig.~\ref{Figure1}). In some networks, links represent associations, e.g., product imports/exports between countries~\cite{NestednessIndustrialEcosystems}, while in others, links represent genuine interactions among system actors, e.g., visitation events between insect pollinators and flowering plants~\cite{BascompteNestedness}. For interaction networks in particular, nestedness has been proposed to affect the feasibility~\cite{Feasibility}, stability~\cite{GhostNestedness,StructuralStability}, robustness~\cite{MarketRobustness}, and persistence~\cite{Persistence} of complex systems.


While the motivations for studying nestedness are grounded in well-established frameworks that provide general explanations for why nested structures are plausible --- e.g., evolutionary theory and trait matching~\cite{PhenotypicComplementarity,EvolProcessesNestedness,EvolSpandrel}, auction theory and loyalty~\cite{MarketRobustness,AuctionTheory} --- less attention has been paid to uncovering the {\it mechanisms} that can generate nestedness. Some process-based models for generating realistic network structures have been proposed, especially in ecology~\cite{PNASrevision1,PNASrevision2,PNASrevision3}, but the focus so far has been on the most appropriate ways of measuring nestedness~\cite{MeasuringNestedness} and, correspondingly, determining whether empirical networks are significantly nested~\cite{ConsumerGuideNestedness}. Various null models have been developed and those that conserve observed degree distributions have been particularly successful at explaining the levels of nestedness seen in empirical networks~\cite{Null1,Null2}. These findings have led to further work demonstrating that skewed degree distributions impose topological constraints that make nested structures likely \cite{BreakingSpellNestedness}, and therefore that the ubiquity of nestedness should be considered unsurprising.


One reason for the difficulty in identifying generative mechanisms is that nestedness has typically been studied in single networks built from time-aggregated data that mask processes occurring at time scales shorter than the aggregation window. However, recent efforts have led to the collection of highly time-resolved interaction data~\cite{FishMarketData, PlantPollData}, which provides an opportunity to explore how the cyclic and seasonal dimensions of social and natural phenomena relate to nestedness at different temporal scales. In ecology, the study of periodic biological events is known as phenology, and various temporal processes have been proposed to explain the overall structure of mutualistic interaction networks~\cite{RampalPhenology,PhenologyCaraDonna,SeasonalDynamics}. Although current explanations for nestedness presuppose elaborate system-specific processes, e.g., trait matching between insect proboscis length and flower corolla depth formed over evolutionary time scales for plant-pollinator networks~\cite{BascompteNestedness}, the availability of time-resolved interaction data means that the effect of more general yet still fundamental processes at shorter time scales, such as how often potential interaction partners overlap through time, can now be investigated. We found that a simple probabilistic model based on phenology could explain and predict the temporal emergence of nested interaction patterns in a broad range of systems. Specifically, the model produced increasingly nested networks as additional temporal data were incorporated, ultimately generating empirically observed levels of nestedness when the amount of temporal aggregation matched a complete phenological cycle.

\section*{Results}



\subsection*{Phenology model}

We analyzed a total of 24 previously published data sets that are publicly available: two of buyer-seller transactions in a fish market in Boulogne-sur-Mer, France~\cite{FishMarketData} and 22 of plant-pollinator visitations sampled across nine countries~\cite{PlantPollData}. All data sets include interactions resolved to the daily level, with data spanning at least one year, or in the case of the ecological data, at least one complete flowering season within a single year (note that data were not collected for all days in an overall sampling period; {\it Methods}). For networks built from the fish market data, nodes represent individual buyers and sellers and links represent at least one transaction of money for goods. One of the two fish market data sets corresponds to an auction market, where buyers and sellers were matched anonymously by an order book; for the other, buyers and sellers could negotiate directly with one another to agree on prices. For the plant-pollinator data, nodes represent species and links represent the observation of at least one visit between individuals belonging to the respective species. To the best of our knowledge, these two groups (fish market and plant-pollinator) of data sets are the only freely available data sets with daily time-resolved interaction data.



For both the fish market and plant-pollinator data sets, interactions (links) could only be realized if system actors (nodes) were physically present together at the same place on a given day; if a particular actor (buyer, seller, pollinator, plant) was not involved in {\it any} interaction on a given day, then we assume the actor was not around that day. This simple yet important temporal feature --- which is shared by most types of interaction networks --- is rarely accounted for when analyzing networks built from aggregated data, despite clearly impacting which interactions are more or less likely to be recorded. To explore this effect on empirical network structures and their levels of nestedness, we developed the {\it phenology model}, which assumes the probability of observing a link increases with the number of times each pair of actors was co-present within a pre-defined aggregation window (Fig.~\ref{Figure1} and {\it Methods}). As our main point of comparison, we used the {\it degree distribution model}, which conserves the degree distributions of empirical networks, since it is the current best explanation for nested structures.



Nestedness was measured using the spectral radius ({\it Methods}), an accurate and reliable approach that correlates strongly with other recommended metrics~\cite{GhostNestedness,MeasuringNestedness}. As the capacity for nested structures is tied to the fill of a network (the fraction of realized links), we scaled nestedness values for networks produced by the phenology and degree distribution models between corresponding values for Erd\"os-R\'enyi random graphs and empirical networks ($\Delta\tilde{\rho}$; {\it Methods}).
\begin{figure*}
\begin{center}
\includegraphics[width=0.43\linewidth]{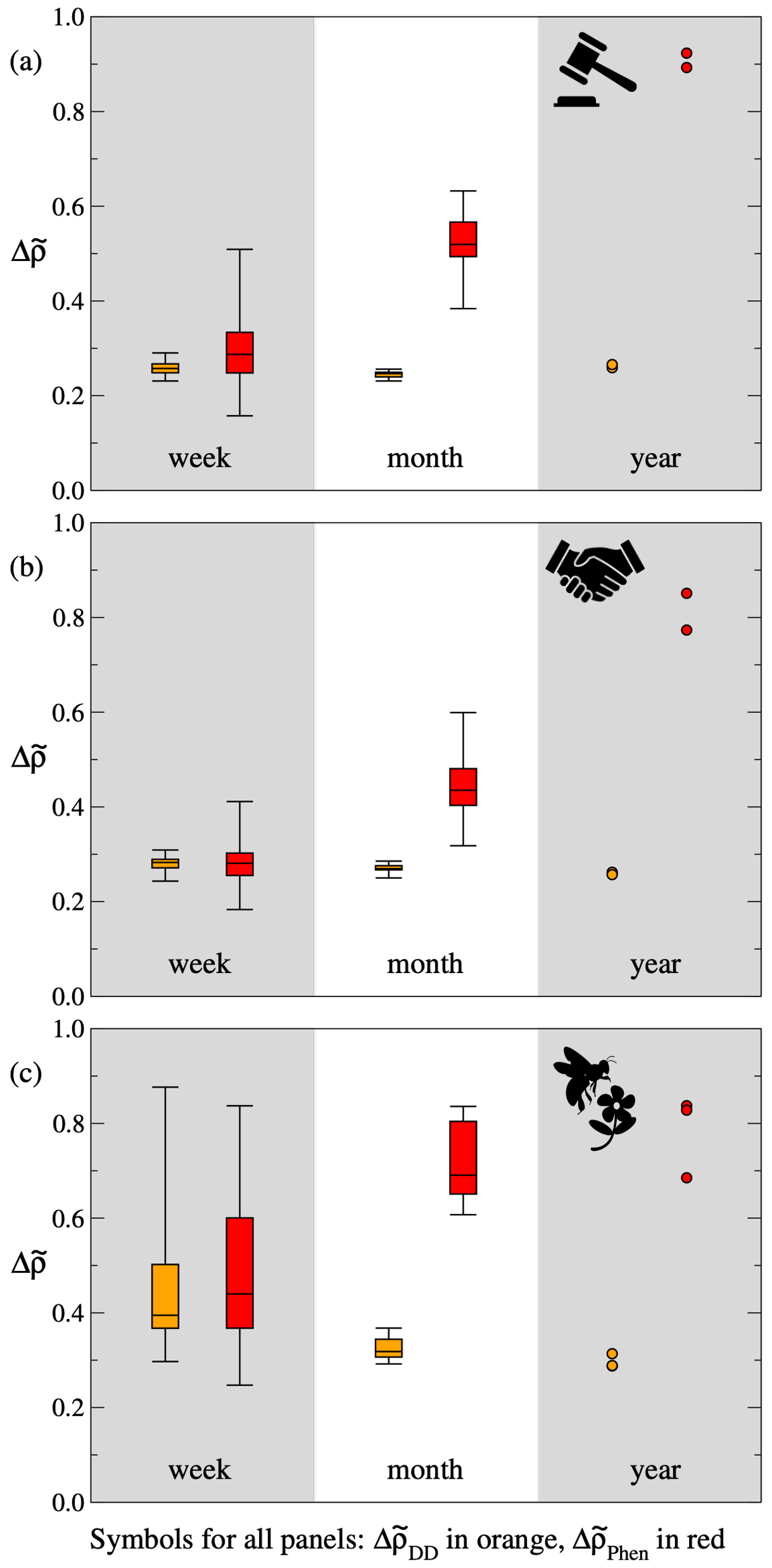}
\end{center}
\caption{Nestedness generated by the degree distribution and phenology models applied to three data sets --- buyer-seller transactions in auction (a) and negotiation (b) fish markets in Boulogne-sur-Mer, France, and plant-pollinator visitations in field sites at Rocky Mountain Biological Laboratory, Colorado, USA (c) --- at three levels of temporal aggregation: week (left), month (middle), and year (right). Values for degree distribution model networks ($\Delta\tilde{\rho}_{\text{DD}}$, orange) and phenology model networks ($\Delta\tilde{\rho}_{\text{Phen}}$, red) are scaled between corresponding values for Erd\"os-R\'enyi random graphs ($\Delta\tilde{\rho}=0$) and empirical networks ($\Delta\tilde{\rho}=1$). Four outlier points have been omitted from the week box plots in panel (c): one for the degree distribution model ($\Delta\tilde{\rho}_{\text{DD}}=1.9$) and three for the phenology model ($\Delta\tilde{\rho}_{\text{Phen}}=1.1,1.2,5.5$). While the levels of nestedness generated by the degree distribution model remain relatively low as temporal aggregation increases, the phenology model generates values that approach empirically observed levels. Full time series for this figure can be found in Supplementary Fig.~S1.
\label{Figure2}}
\end{figure*}



\subsection*{Emergence of nestedness}

At all levels of temporal aggregation (week, month, year), empirical networks were more nested than Erd\"os-R\'enyi random graphs, as were networks produced by the degree distribution and phenology models (Fig.~\ref{Figure2}). Interestingly, when daily interaction data were aggregated to weekly networks, phenology and degree distribution models generated similar levels of nestedness, around a quarter of that for empirical networks. For monthly aggregated networks, the proportion of generated nestedness rose to roughly half for the phenology model, but remained at around a quarter for the degree distribution model. Not all of the 22 plant-pollinator data sets were suitable for constructing weekly and monthly networks (Supplementary Table~S1), but in all cases we could reliably analyze data aggregated over an entire year, i.e., a single field season --- the usual temporal resolution of published plant-pollinator networks. At the annual level, across all 24 data sets, the degree distribution model generated around 30\% of empirically observed nestedness, compared to around 80\% for the phenology model (Table~\ref{Table1}). Thus, the proportion of nestedness explained by phenology increases as temporal data are aggregated, reaching close to empirically observed values as the level of aggregation matches a complete phenological cycle.
\begin{table*}
\caption{Empirical network properties and proportion of empirical nestedness generated by the degree distribution ($\Delta\tilde{\rho}_{\text{DD}}$) and phenology ($\Delta\tilde{\rho}_{\text{Phen}}$) models at an annual level of temporal aggregation, mean and standard deviation (in parentheses) of 1000 realizations.
\label{Table1}}
\begin{center}
\begin{tabular}{llrrrrrll}
Data set & Ref & Rows & Cols & Fill & Days & Years & $\Delta\tilde{\rho}_{\text{DD}}$     & $\Delta\tilde{\rho}_{\text{Phen}}$    \\
Auction &\cite{FishMarketData}        & 195  & 100  & 0.38 & 525  & 2     & 0.26 (0.02) & 0.91 (0.02) \\
Negotiation & \cite{FishMarketData}    & 207  & 93   & 0.50 & 538  & 2     & 0.26 (0.03) & 0.81 (0.03) \\
CaraDonna     & \cite{CaraDonnaData} & 46   & 93   & 0.14 & 106  & 3     & 0.30 (0.09) & 0.78 (0.10) \\
Alarcon2008   & \cite{Alarcon2008} & 56   & 232  & 0.07 & 50   & 3     & 0.36 (0.08) & 0.63 (0.08) \\
Benadi2014    & \cite{Benadi2014} & 121  & 406  & 0.04 & 46   & 1     & 0.27 (0.02) & 0.83 (0.03) \\
Burkle2012    & \cite{Burkle2012} & 45   & 127  & 0.08 & 17   & 1     & 0.28 (0.05) & 0.54 (0.05) \\
Chacoff2018   & \cite{Chacoff2018} & 59   & 198  & 0.09 & 93   & 6     & 0.31 (0.09) & 0.73 (0.10) \\
Fruend2010    & \cite{Fruend2010} & 103  & 160  & 0.06 & 51   & 1     & 0.29 (0.05) & 1.03 (0.06) \\
Lara-Romero2016& \cite{Lara-Romero2016} & 22   & 146  & 0.16 & 22   & 1     & 0.30 (0.06) & 0.82 (0.07) \\
LeBuhnYY      & \cite{PlantPollData} & 90   & 134  & 0.05 & 74   & 2     & 0.29 (0.07) & 0.89 (0.08) \\
Olito2015     & \cite{Olito2015} & 43   & 125  & 0.06 & 32   & 1     & 0.33 (0.08) & 1.02 (0.09) \\
Rasmussen2013 & \cite{Rasmussen2013} & 39   & 108  & 0.10 & 106  & 2     & 0.31 (0.09) & 1.01 (0.11) \\
ResascoYY     & \cite{PlantPollData} & 35   & 208  & 0.08 & 51   & 3     & 0.37 (0.10) & 0.71 (0.11) \\
Simanonok2014 & \cite{Simanonok2014} & 45   & 126  & 0.07 & 44   & 1     & 0.29 (0.06) & 0.93 (0.07) \\
Thompson2018  & \cite{Thompson2018} &  41   & 127  & 0.09 & 37   & 1     & 0.35 (0.07) & 0.99 (0.08) \\
Vazquez2003   & \cite{Vazquez2003} & 14   & 117  & 0.13 & 76   & 1     & 0.44 (0.13) & 0.92 (0.12) \\
Weiner2014    & \cite{Weiner2014} & 97   & 405  & 0.05 & 70   & 1     & 0.29 (0.02) & 0.81 (0.02) \\
Winfree2014   & \cite{Winfree2014} & 59   & 85   & 0.09 & 33   & 2     & 0.31 (0.10) & 0.67 (0.11) \\
WinfreeYYa    & \cite{WinfreeYYa} & 113  & 120  & 0.08 & 77   & 3     & 0.30 (0.08) & 0.82 (0.09) \\
WinfreeYYb    & \cite{WinfreeYYa} & 112  & 143  & 0.07 & 81   & 3     & 0.29 (0.06) & 0.83 (0.07) \\
WinfreeYYc    & \cite{PlantPollData} & 14   & 33   & 0.21 & 8    & 1     & 0.35 (0.14) & 0.50 (0.14) \\
WinfreeYYd    & \cite{WinfreeYYd} & 101  & 95   & 0.12 & 54   & 1     & 0.26 (0.03) & 0.80 (0.03) \\
WinfreeYYe    & \cite{PlantPollData} & 84   & 79   & 0.12 & 72   & 3     & 0.30 (0.07) & 0.96 (0.10) \\
WinfreeYYf    & \cite{WinfreeYYf} & 57   & 84   & 0.08 & 22   & 1     & 0.27 (0.06) & 0.54 (0.07)
\end{tabular}
\end{center}
\end{table*}




In order to see how closely model predictions matched empirical data, we used an assessment scheme that classified each link prediction into one of four categories: true positive, true negative, false positive, false negative (Fig.~\ref{Figure1}). From this information, we measured model accuracy using two summary statistics: the phi coefficient and the F-score ({\it Methods}). The phi coefficient is the equivalent of the Pearson correlation coefficient but for two binary variables rather than two continuous variables; for interaction networks, the binary variables are the presence or absence of each potential link in a model-generated network compared to its presence or absence in the empirical network. The phi coefficient provides a full-network assessment of model accuracy and is complemented by the F-score, which focuses on the subset of realized links (i.e., ``positive'' observations) in the empirical network. In the context of the phenology model, the F-score measures the relative contributions of actor co-presences (i.e., true positives) versus interaction preferences (departures from co-presence, i.e., false positives and false negatives) to network structure.

The phenology model outperformed the degree distribution model for all 24 data sets (Supplementary Table~S2). Additionally, the amount of nestedness ($\Delta\tilde{\rho}$) generated by the phenology model was strongly correlated with the phi coefficient --- Spearman's rank correlation coefficient, $r=0.42$, $p=0.04$, $n=24$ --- and to a lesser extent the F-score, $r=0.29$, $p=0.17$, $n=24$. By contrast, for the degree distribution model, there was no correlation between nestedness and the phi coefficient, $r=-0.05$, $p=0.82$, $n=24$, nor nestedness and the F-score, $r=0.03$, $p=0.88$, $n=24$. For the phenology model, at the annual level of temporal aggregation (the level at which all data sets could be compared), there was no notable difference in phi coefficients between fish market and plant-pollinator data sets. However, F-scores for the two fish market data sets, $F_1=0.67$ for the auction market and $F_1=0.68$ for the negotiation market, were much higher than the average value for the 22 plant-pollinator data sets, $F_1=0.33\pm0.08$ (mean~$\pm$~standard deviation). For the annual plant-pollinator networks, this F-score result suggests that roughly one-third of links can be explained by the relative availability of interaction partners. The remaining two-thirds of links, which represent observed interactions between plants and pollinators that are rarely co-present (i.e., phenology model false negatives), may then be attributed either to opportunistic interactions or to preferences for particular interaction partners that are not routinely available.


\subsection*{Backbone of nested interactions}

Interestingly, the links that were most readily explained by frequent actor co-presences appear to form a {\it backbone of nested interactions}. We investigated this possibility by comparing the relative levels of nestedness for the subset of true positive links (attributable to co-presences) to the subset of false negative links (attributable to interaction preferences). For 15 of the 24 data sets, we found that: (i) true positive links were significantly more nested than false negative links, and (ii) true positive links were significantly more nested relative to Erd\"os-R\'enyi random graphs than false negative links were to Erd\"os-R\'enyi random graphs (Supplementary Table~S3). This finding indicates that the set of interactions generated by frequent actor co-presences is more nested than the other, remaining set of interactions, which are distributed more randomly among actors.



\section*{Discussion}

\subsection*{Importance of time}

In general, system dynamics play an integral role in structuring interaction networks. This influence ranges from straightforward and unambiguous, e.g., the impossibility of an interaction if two actors are never around at the same time, to complex and subtle, e.g., interaction partners co-evolving in ways that alter their ability to interact with other actors. Here, we explored the relationship between nestedness and how often actors overlap in space and time. We proposed the phenology model, which assumes the more often actors overlap the more likely they are to interact, and applied it to buyer-seller fish market networks and plant-pollinator visitation networks. In this way, we sought to understand nestedness in terms of potential generative mechanisms --- temporal processes that drive the emergence of nestedness in empirical data --- rather than topological properties that are known only {\it after} networks have been constructed.

Using daily interaction data, we found that the proportion of empirically observed nestedness generated by the phenology model increased with the level of temporal aggregation used to construct networks. By contrast, the performance of the degree distribution model remained relatively poor as the data aggregation window increased from week to month to year. A formal model selection procedure is difficult to specify due the different conceptual motivations for the two models and the different types of data required to parameterize them. For the degree distribution model, link probabilities are derived from the topology of a final time-aggregated network, whereas for the phenology model, link probabilities are based on the number of times actors are co-present within a given sampling period. For the data sets analyzed here, it is clear that the phenology model performed noticeably better than the degree distribution model at time scales longer than a week; nevertheless, in future studies it will still be worth considering both models (or variants of the two models) in parallel.

The closest agreement between the phenology model and empirical data was for annual networks, for which the model was also able to predict a large fraction of empirical links. There are both technical and scientific reasons for this result. For weekly networks, there are inherently little data for the phenology model to work with and therefore limited opportunities for it to produce network structures that differ significantly from Erd\"os-R\'enyi random graphs. At such short time scales, with the plant-pollinator data in particular, the phenology model is mostly picking up signals related to sampling effort and relative abundance, namely, that actor presences and absences in the model reflect high and low species abundances more so than phenological effects based on reliable estimates of co-presence. For monthly networks, the maximum number of possible co-presences increases from seven to 31 days, thereby providing more leeway to test the effect of co-presence frequency on network structure. For annual networks, data aggregation covers a complete phenological cycle (e.g., one financial year, one full flowering season) and offers the widest range of possible and recorded co-presence values.

In histograms of the relative frequency of co-presences at the three levels of temporal aggregation, while the median number of co-presences hardly increases, e.g., auction market: $1.1\pm0.3$ for weeks (mean~$\pm$~standard deviation for 91 networks), $2.3\pm0.5$ for months (23 networks), and $8.5\pm0.7$ for years (2 networks), the right tail of the distribution becomes noticeably fatter, e.g., auction market largest co-presence values: $4.1\pm1.0$ for weeks, $13.2\pm2.6$ for months, and $117.5\pm0.7$ for years (Supplementary Figs.~S2--S4). The increasing weight in the right tail of the co-presence distribution allows the phenology model to pick out a more consistent set of (nested) interactions. This analysis reiterates the importance of matching the level of temporal aggregation to the natural phenological time scale of a system to maximize the predictive power of simple temporal models like the phenology model.

\subsection*{Presence versus co-presence}

The phenology model assumes the more often two actors are co-present the more likely they are to interact. It is reasonable to expect some actors are frequently co-present with other actors simply because they are generally more present during a given time period. Since it is easier to assess the relative presences of actors compared to collecting additional information on how often actors overlap through time, it is interesting to see how well presence, rather than co-presence, explains empirical nestedness. To do so, we designed the {\it presence model} by modifying the phenology model slightly: instead of scaling the probability of an interaction by the number of days two actors overlap, we scale the probability by the average number of days the actors were present during a specified time period ({\it Methods}).

The proportion of empirical nestedness generated by the presence model typically fell between values for the degree distribution and phenology models but was higher for the auction market data set and 6 of the 22 plant-pollinator data sets (Supplementary Table~S4). At the annual level, across all 24 data sets, the presence model generated around 70\% of empirically observed nestedness, compared to 30\% and 80\% for the degree distribution and phenology models, respectively. While the presence model generated relatively high nestedness for the auction market data set ($\Delta\tilde{\rho}=1.04$), the value was much lower for the negotiation market data set ($\Delta\tilde{\rho}=0.37$). However, link prediction accuracies were similar for the two fish market data sets, around 64\% of the explanatory power of the phenology model for the phi coefficient and 88\% for the F-score. This slightly worse performance extended to the plant-pollinator data sets, with the presence model offering around 80\% of the explanatory power of the phenology model for the phi coefficient and 87\% for the F-score (Supplementary Table~S4).

Interestingly, while the proportion of empirical nestedness generated by the presence model differed notably between the two fish market data sets, prediction accuracies were similar. The lower nestedness ($\Delta\tilde{\rho}$) for the negotiation market data set suggests that predictions were concentrated on links between pairs of high-presence actors with similar numbers of interaction partners, at the expense of predictions involving high-presence actors with few interaction partners. On the other hand, the presence model generated higher nestedness for the auction market data set because, in addition to predicting interactions among a core of high-presence, high-degree actors, the model also predicted interactions between high-presence, low-degree actors and high-presence, high-degree actors (likely market makers). The fact that prediction accuracy was similar between auction and negotiation market data sets suggests that correct predictions were mainly restricted to links among a core set of high-presence, high-degree actors.

These findings extend our understanding of how co-occurrence patterns inform interaction patterns~\cite{NEXUSrevision1}. In particular, compared to the presence model, the phenology model improves link prediction accuracy on two fronts by: (i) refining which potential interactions among high-presence, high-degree actors are most likely to be realized based on how often they overlap through time and (ii) enabling more accurate predictions involving low-degree actors by accounting for how often they are co-present with high-degree actors, and not just how often they are present.

\subsection*{Future directions}

Here, information on interaction partner availability through time was inferred from empirical data and implemented in the phenology model as simply the number of co-presences between each pair of actors. In future work, it will be worth exploring in detail how phenologies combine to produce nested structures. Accounting for exactly when and for how long actors are active is especially valuable for understanding the interactions of infrequent participants which, by definition, are associated with low co-presence values. These interactions are likely classified as false negatives, but considering phenologies more fully would help distinguish, for example, the genuine interactions of a rare plant with a long flowering period that is visited by a large number of short-lived pollinator species from the background noise of actor pairs that barely overlap.

The phenology model presents a viable path to a general mechanistic understanding of interaction network structure. Currently, the model explains the formation of backbones of nested interactions comprising around two-thirds of links in fish market networks and one-third of links in plant-pollinator networks; but this still leaves a large number of links that cannot be explained by actor co-presence frequency. Since the model assigns distinct probabilities to each potential link, systematically modifying these probabilities is one way of incorporating additional mechanisms. For plant-pollinator networks, this could include the effects of competition~\cite{ValdovinosAdaptiveForaging}, phenotypic complementarity~\cite{PhenotypicComplementarity,TraitBasedNetworkAssembly}, and interaction preferences~\cite{InteractionPrefs}. For market-based networks, especially those in which buyers and sellers know one another and repeated transactions are common, loyalty is an prominent feature that impacts which interactions are likely to be realized~\cite{MarketRobustness,FishMarketData}. Interestingly, for the fish market data, auction networks were slightly more nested than negotiation networks. The phenology model also generated auction networks that were relatively more nested than negotiation networks, suggesting that co-presence had a greater impact on nestedness when there was automatic matching of buyers and sellers through an order book (auction market) compared to when buyers and sellers made deals directly (negotiation market).

Since nestedness co-varies with other network metrics~\cite{CorrelationNestednessMetrics}, it would be interesting to measure the extent to which an updated phenology model could explain the emergence of --- and trade-offs between --- other structural patterns, such as modularity and community structure~\cite{Modularitynestedness,CommunityDetection}.
The fish market and plant-pollinator data sets both describe positive and mutually beneficial interactions between actors. It would be useful to contrast findings with data sets that include negative interactions, such as competition networks in which actors are negatively affected by the presence of other actors, or food webs in which one actor (a predator) benefits at the expense of another actor (a prey).

Although we have focused on binary networks, interaction partner availability and higher-level selection processes will also have an effect on the structure of quantitative or weighted networks that additionally document the relative strength or frequency of interactions~\cite{NestednessReview}. Indeed, it may be even more important to account for actor timing when dealing with weighted interaction networks. For ecological systems, for example, phenology often dictates the abundance of species through regular growth and reproduction cycles, with the relative abundance of species, in turn, known to impact weighted network structure~\cite{GhostNestedness,PhenologyCaraDonna}. Finally, it is worth noting that although the current study is focused on the temporal origin of nestedness in established and stable systems over relatively short and cyclical time frames, the phenology model would also be useful for studying interaction patterns as markets form and collapse and ecological communities assemble and disassemble~\cite{PNASrevision4,Commassemblynetworks,Assemblingmutualisticcommunities}.

\section*{Concluding remarks}

For interaction networks, some influence of time --- and of timing --- on network structure is to be expected. A buyer and seller must agree on a price to make a trade, and a plant must be flowering for a pollinator to visit it. However, these short-term temporal dependencies are obscured in the time-aggregations commonly applied in traditional network analyses. In the case of plant-pollinator networks, this seemed acceptable since nestedness was thought to reflect constraints on trait matching shaped over evolutionary time scales. But we have now shown that phenology, which concerns much shorter, ecological time scales, can also lead to nested structures. It will therefore be important to revisit the relationship between nestedness and system-level properties such as feasibility, stability, robustness, and persistence, given the current uncertainty in how the two, temporally disconnected processes combine to generate nestedness and other structural features. For ecological interactions, anthropogenic drivers like climate change and urbanization are making phenological mismatches more likely, meaning there is an urgent need to trace the effects of shifting phenologies on network structure and the resulting impacts on species fitness and survival.

\section*{Materials and methods}

\noindent \textbf{Data.} Auction and negotiation market data sets were obtained from an earlier study~\cite{FishMarketData}. Data on daily buyer-seller transactions were collected over 21 months spanning two consecutive calendar years; we divided it into two parts comprising 10.5 months each, representing two ``years'' of daily buyer-seller transaction data. The 22 plant-pollinator data sets were previously compiled as part of an earlier study~\cite{PlantPollData} and were selected for analysis here because they include visitations between individual insect pollinators and plants resolved to the daily level; source papers are listed in Table~\ref{Table1}. Note that plant-pollinator data were not collected for all days in the overall sampling period of each data set; similarly, buyer-seller data excluded days when the market was closed, e.g., Sundays and national holidays. Temporally aggregated networks were built from consecutive, non-overlapping blocks of daily interaction data and only included nodes and links corresponding to buyer and seller individuals or plant and pollinator species that interacted during a given block of time. Multiple interactions in a single block were counted only once, leading to binary, interaction presence-absence networks.

\vspace{2mm}
\noindent \textbf{Models.} We considered four probabilistic models that conserve, on average, the total number of links in a time-aggregated empirical network. The Erd\"os-R\'enyi random graph and degree distribution model are well-established, while the phenology and presence models are newly presented here. Each model describes links as independent Bernoulli random variables with probabilities defined below.

\vspace{2mm}
\noindent {\it Erd\"os-R\'enyi random graph.} Each link has the same probability of being realized, $p^{\text{ER}}_{ij} = L/(RC)$ (the fill or connectance of an empirical network); where $L$ is the number of links in an empirical network, $R$ is the number of nodes in the actor category listed as rows, e.g., sellers or plants, $C$ is the number of nodes in the actor category listed as columns, e.g., buyers or pollinators, and $i$ and $j$ index nodes in the row category and column category, respectively.

\vspace{2mm}
\noindent {\it Degree distribution model.} Link probability is $p^{\text{DD}}_{ij} = k_i/(2C) + k_j/(2R)$; where $k_i$ is the empirical degree of node $i$ in the row category and $k_j$ is the empirical degree of node $j$ in the column category. In addition to the total number of links in an empirical network, the degree distribution model also conserves the expected number of links associated with each node and therefore the expected degree distribution of the empirical network~\cite{ConsumerGuideNestedness}.

\vspace{2mm}
\noindent {\it Phenology model.} The model requires two parameters that are not derived from the topology of an empirical network. First is the sampling period, $f$, which is the number of days worth of interaction data included in the time-aggregated empirical network (if a different minimal temporal unit for resolving interaction data is used then $f$ is specified in terms of this unit, e.g., hours instead of days). The second parameter is the co-presence matrix, $N$, that lists the number of instances --- in terms of days or some other minimal temporal unit --- each pair of actors overlap in space and time during the time-aggregated sampling period. If the probability of an interaction given a co-presence is $p$ and the probability of no interaction is $q = 1 - p$, then the probability of at least one interaction during the sampling period and therefore the probability of a link is $p^{\text{Phen}}_{ij} = 1 - q^{n_{ij}}$; where $n_{ij}$ is the number of co-presences between actors represented by nodes $i$ and $j$. In this way, the greater the number of co-presences, the higher the probability of a link.

\vspace{2mm}
\noindent Note that we assume a single probability of an interaction given a co-presence (i.e., $p$), which we set to a value such that the expected number of links equals the number of links in the empirical network. Since there are at most $f + 1$ distinct entries in $N$, ranging from $0$ (i.e., actors never overlap during the sampling period) to $f$ co-presences (i.e., actors always overlap), we find $p = 1 - q$ by solving the polynomial, $\sum^{k=f}_{k=0} x_k q^k - Z = 0$; where $x_0$ is the number of $0$s in $N$, $x_1$ the number of $1$s, etc., and $Z = RC - L$ is the number of zeros in the incidence matrix representation of the empirical network.

\vspace{2mm}
\noindent For the example in Fig.~\ref{Figure1}, $f = 6$ days and the co-presence matrix, $N$, is shown in panel b. Solving the corresponding polynomial, $q^0 + 2q + 4q^2 + 2q^3 - 3 = 0$, gives
$q = 0.47$ and $p = 1 - 0.47 = 0.53$. Individual link probabilities in panel c can then be calculated from these values of $p$ and $q$.

\vspace{2mm}
\noindent {\it Presence model.} Based on the phenology model, the presence model is designed to provide a comparison of the effect of actor presence on network structure relative to the effect of actor co-presence. Instead of the number of days each pair of actors overlap, the average number of days each pair of actors is present during the time-aggregated sampling period is listed in matrix $N$ (in practice, the average number of days is rounded up to the nearest integer to ensure the polynomial described above is solvable). Everything else remains the same as in the phenology model. Note, however, that the interpretation of $p$ as the probability of an interaction given a co-presence no longer holds, and therefore the presence model should be seen as a way to scale link probabilities according to actor prevalence or commonness, i.e., presence, rather than opportunities to interact, i.e., co-presence.

\vspace{2mm}
\noindent \textbf{Measuring nestedness.} A network can be represented as an adjacency matrix in which the full set of nodes is listed as both rows and columns and non-zero entries indicate the presence of a link between two nodes. For bipartite networks, there will be two blocks of zeros on the diagonal since there are no links between nodes in the same category; the (symmetric) off-diagonal block contains all relevant topological information and is known as the incidence matrix, as in the upper box of Fig.~\ref{Figure1}. Measuring nestedness involves computing the spectral radius --- the absolute value of the dominant or largest eigenvalue --- of the adjacency matrix~\cite{GhostNestedness,MeasuringNestedness,SpectralRadiusMath}. For a given size and fill, i.e., total number of nodes and links, the larger the spectral radius the more nested the particular network topology. To enable comparison between different networks, raw values of the spectral radius can be normalized by their maximum possible values~\cite{rhoUpperBound} to give a measure for nestedness that ranges between $\sim0$ and $1$: $\rho = \rho_{\text{raw}} / \sqrt{2L - R - C + 1}$.

\vspace{2mm}
\noindent We used normalized spectral radius values of Erd\"os-R\'enyi random graphs as a baseline for contextualizing the relative nestedness of empirical networks and more complex model-generated networks (here, the degree distribution and phenology models): $\Delta\rho = \rho - \rho_{\text{ER}}$; where $\rho_{\text{ER}}$ is the mean value of the normalized spectral radius of an ensemble of Erd\"os-R\'enyi random graphs (here, we used 10,000 networks) parameterized according to the the number of nodes and links in the focal empirical network. Empirical or model-generated networks with $\Delta\rho > 0$ are nested, while networks with $\Delta\rho < 0$ are anti-nested~\cite{GhostNestedness}. To more easily compare models and their ability to generate empirically observed levels of nestedness, we considered a further rescaling: $\Delta\tilde{\rho} = (\rho - \rho_{\text{ER}})/(\rho_{\text{emp}} - \rho_{\text{ER}})$; where $\rho_{\text{emp}}$ is the normalized spectral radius of the focal empirical network. A model-generated network with $\Delta\tilde{\rho} = 0$ has the same level of nestedness as a corresponding Erd\"os-R\'enyi random graph, while $\Delta\tilde{\rho} = 1$ indicates that the model-generated network has the same level of nestedness as the empirical network.

\vspace{2mm}
\noindent \textbf{Assessing link prediction accuracy.} We used the confusion matrix, a two dimensional contingency table, to assess the accuracy of model predictions~\cite{ConfusionMatrix}. For a given model-generated network, each entry in its corresponding incidence matrix can be assigned to one of four categories: true positive (TP) --- link present in both model and empirical networks; true negative (TN) --- link absent in both model and empirical networks; false positive (FP) --- link present in model network but absent in empirical network; false negative (FN) --- link absent in model network but present in empirical network. We summarized model accuracy using two well-established statistics: the phi coefficient and the F-score~\cite{PhiFscore}.

\vspace{2mm}
\noindent {\it phi coefficient.} This statistic involves all four categories and is the equivalent of the Pearson correlation coefficient for two binary variables: $\phi = \frac{\text{TP} \times \text{TN} - \text{FP} \times \text{FN}}{\sqrt{(\text{TP} + \text{FP})(\text{TP} + \text{FN})(\text{TN} + \text{FP})(\text{TN} + \text{FN})}}$. Values for the phi coefficient range from $-1$ (perfect disagreement) through $0$ (no relationship) to $1$ (perfect agreement). Note that since we constrain model-generated networks to have, on average, the same number of links as in an empirical network, FP $\approx$ FN and therefore phi coefficient values in this case approximate values for the informedness or Youden's J statistic, $J = \text{sensitivity} + \text{specificity} - 1 = \text{TP} / (\text{TP} + \text{FN}) + \text{TN} / (\text{TN} + \text{FP}) - 1$.

\vspace{2mm}
\noindent {\it F-score.} This statistic focuses on the subset of realized links in an empirical network and as such does not involve true negatives: $F_1 = 2\text{TP} / (2\text{TP} + \text{FP} + \text{FN})$. Values for the F-score range from $0$ (no links correctly predicted) to $1$ (all links correctly predicted). Since FP $\approx$ FN, F-score values in this case approximate values for the sensitivity or true positive rate, $\text{TPR} = \text{TP} / (\text{TP} + \text{FN})$. It is worth noting that, for Erd\"os-R\'enyi random graphs, sensitivity values (and therefore F-score values, here) approximate the fill or connectance, $L/RC$, of an empirical network.

\vspace{2mm}
\noindent \textbf{Backbone of nested interactions.} We tested the hypothesis that the subset of true positive links predicted by the phenology model are more nested than the subset of false negative interactions in two parts. First, we calculated the ratio of nestedness values for the subsets of true positive and false negative links; if $\rho_{\text{TP}}/\rho_{\text{FN}} > 1$, then interactions between frequently co-present actors are more nested than the remaining interactions. To account for any confounding size-based effects, we also compared the deviations of nestedness from comparable Erd\"os-R\'enyi random graphs. Specifically, we calculated the proportion of excess nestedness for the phenology model relative to the maximum possible value: $\rho_{\text{TP},\Delta\text{ER}} = (\rho_{\text{TP}} - \rho_{\text{TP,ER}})/(1 - \rho_{\text{TP, ER}})$ for the subset of true positive links and $\rho_{\text{FN,ER}} = (\rho_{\text{FN}} - \rho_{\text{FN,ER}})/(1 - \rho_{\text{FN,ER}})$ for the subset of false negative links. If $\rho_{\text{TP},\Delta\text{ER}} / \rho_{\text{FN},\Delta\text{ER}} > 1$, then interactions between frequently co-present actors are more nested relative to Erd\"os-R\'enyi random graphs than the remaining interactions.

\vspace{5mm}
\noindent \textbf{Data availability.} All data analyzed in this study have been published previously and are available through the references in Table~1. Code to perform analyses are available on GitHub in the public repository, \url{https://github.com/pstaniczenko/TemporalOriginNestedness}.

\vspace{2mm}
\noindent \textbf{Author contributions.} P.P.A.S. and D.P. designed research, performed research, analyzed data, and wrote the paper; P.P.A.S. also contributed new analytical tools.

\vspace{2mm}
\noindent \textbf{Funding statement.} There were no funders for this work.

\vspace{3mm}

\bibliographystyle{Science}
\bibliography{literature.bib}

\end{document}


\maketitle

\noindent \textbf{Table \ref{TableS1}.} Proportion of empirical nestedness generated by the degree distribution and phenology models at three levels of temporal aggregation: weekly, monthly, and annual.

\vspace{6pt}

\noindent \textbf{Table \ref{TableS2}.} Phi coefficients and F-scores for the degree distribution and phenology models at three levels of temporal aggregation: weekly, monthly, and annual.

\vspace{6pt}

\noindent \textbf{Table \ref{TableS3}.} Backbone of nestedness results for the phenology model at an annual level of temporal aggregation.

\vspace{6pt}

\noindent \textbf{Table \ref{TableS4}.} Proportion of empirical nestedness, phi coefficients, and F-scores for the degree distribution, presence, and phenology models at an annual level of temporal aggregation.

\vspace{6pt}

\begin{figure*}[ht]
\begin{center}
\includegraphics[width=1\linewidth]{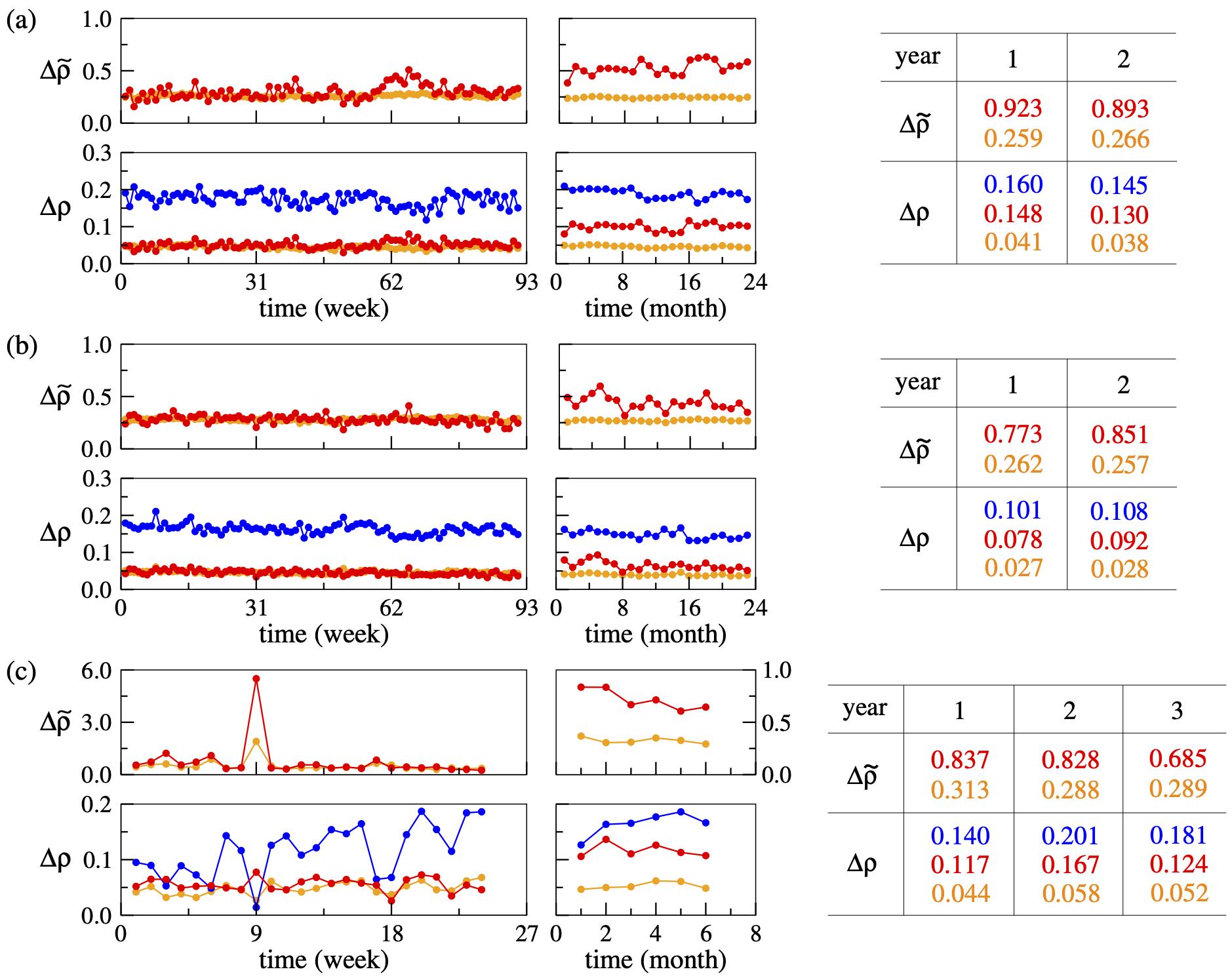}
\end{center}
\caption{Nestedness generated by the degree distribution and phenology models relative to empirical networks built from three data sets --- buyer-seller transactions in auction (a) and negotiation (b) fish markets in Boulogne-sur-Mer, France, and plant-pollinator visitations in field sites at Rocky Mountain Biological Laboratory, Colorado, USA (c) --- at three levels of temporal aggregation: week (left), month (middle), and year (right). In the lower plots of each panel, nestedness values for empirical networks (blue), degree distribution model networks (orange), and phenology model networks (red) are measured relative to corresponding values for Erd\"os-R\'enyi random graphs ($\Delta\rho=0$); in the upper plots, nestedness values for the two models are scaled between corresponding values for Erd\"os-R\'enyi random graphs ($\Delta\tilde{\rho}=0$) and empirical networks ($\Delta\tilde{\rho}=1$). While the levels of nestedness generated by the degree distribution model remain relatively low as temporal aggregation increases, the phenology model generates values that approach empirically observed levels.
\label{FigureS1}}
\end{figure*}

\noindent \textbf{Figure \ref{FigureS1}.} Version of Figure 2 in the main paper with time series of nestedness values.

\vspace{6pt}

\begin{figure*}[ht]
\begin{center}
\includegraphics[width=1\linewidth]{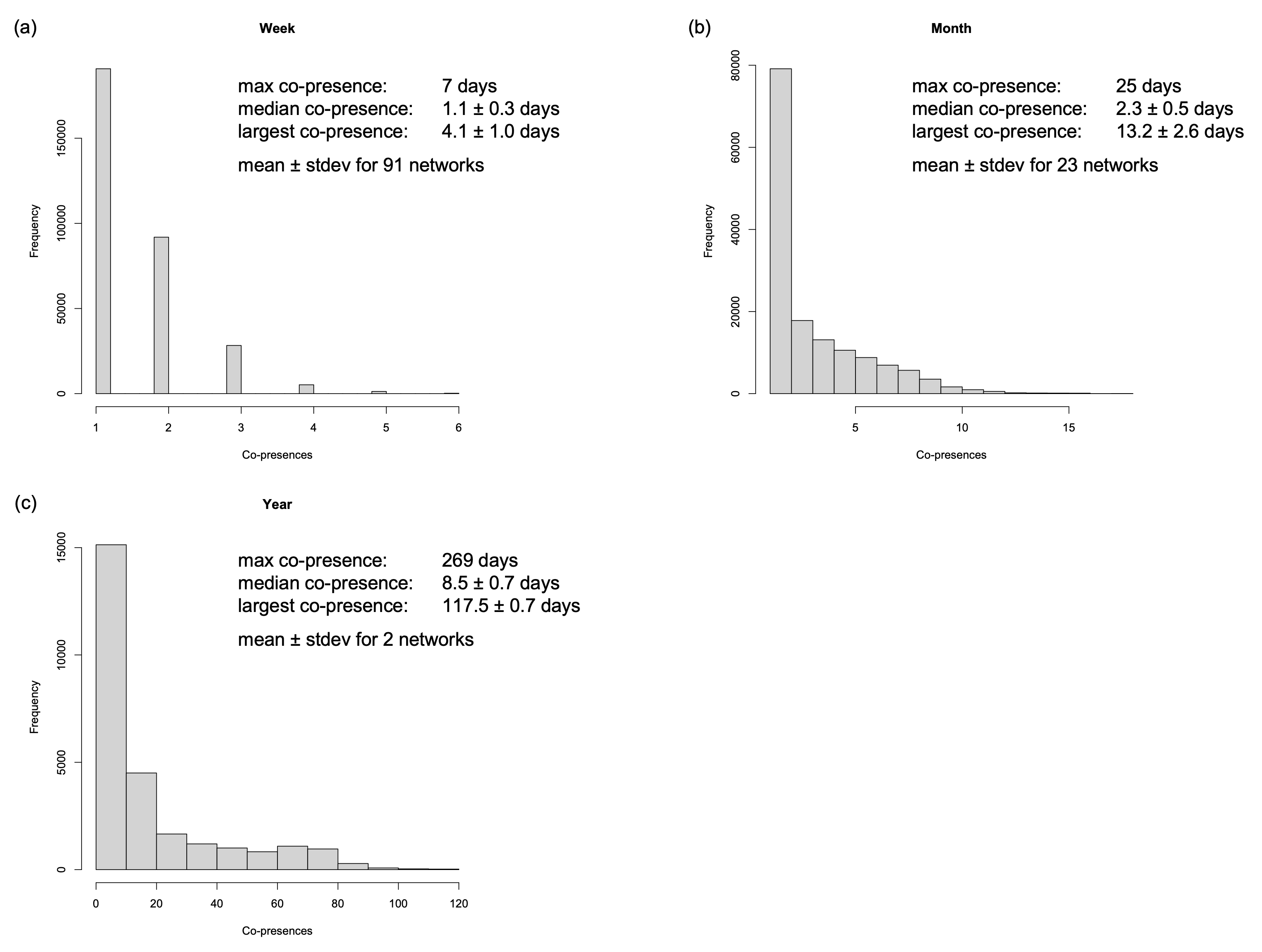}
\end{center}
\caption{Histograms of the relative frequency of co-presences for auction market data~\cite{FishMarketData}. Panels show co-presences for all networks combined at the specific level of temporal aggregation while statistics for the median and largest number of co-presences statistics present averages and standard deviations for individual networks.
\label{FigureS2}}
\end{figure*}

\noindent \textbf{Figure \ref{FigureS2}.} Histograms of the relative frequency of co-presences for auction market data.

\vspace{6pt}

\begin{figure*}[ht]
\begin{center}
\includegraphics[width=1\linewidth]{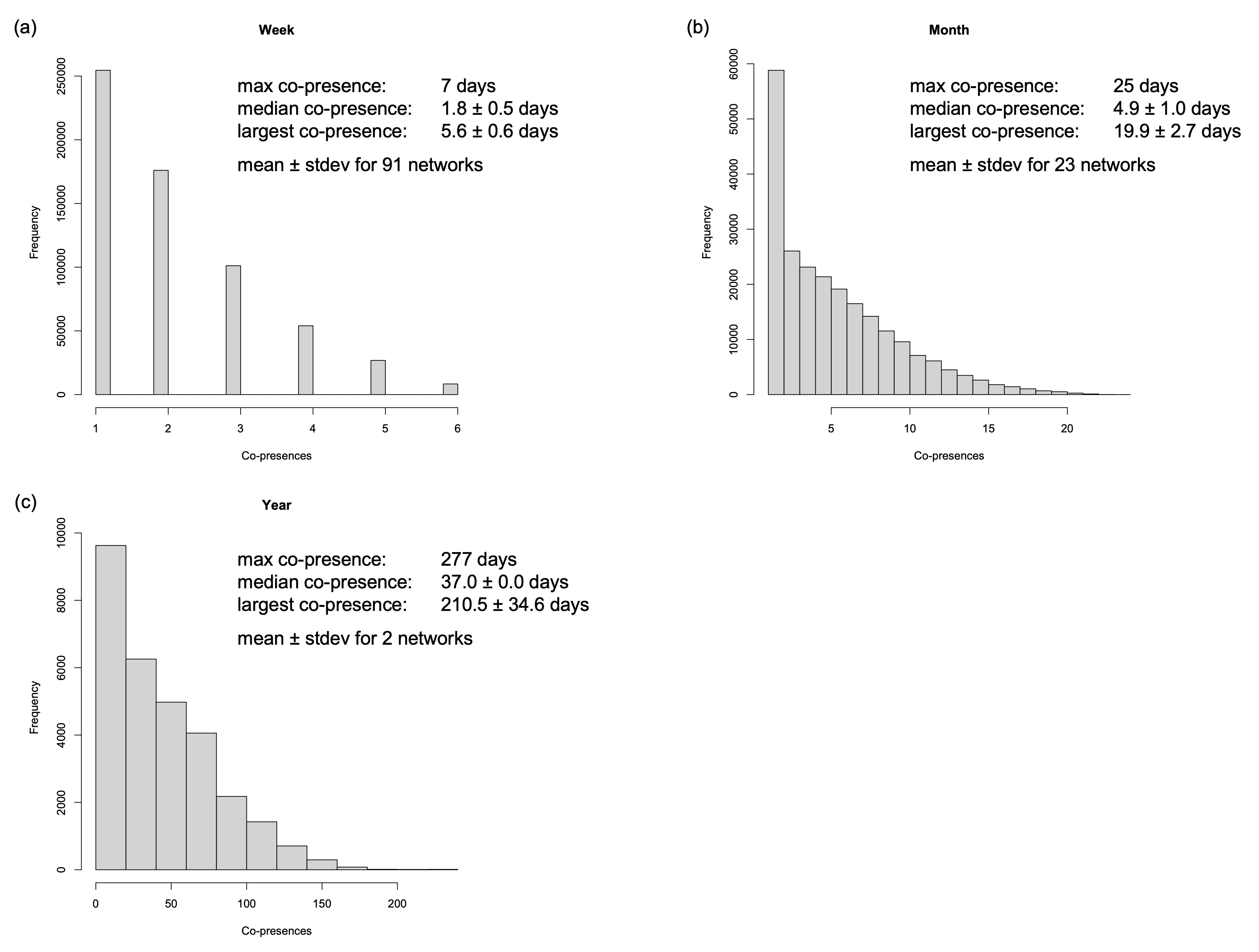}
\end{center}
\caption{Histograms of the relative frequency of co-presences for negotiation market data~\cite{FishMarketData}. Panels show co-presences for all networks combined at the specific level of temporal aggregation while statistics for the median and largest number of co-presences statistics present averages and standard deviations for individual networks.
\label{FigureS3}}
\end{figure*}

\noindent \textbf{Figure \ref{FigureS3}.} Histograms of the relative frequency of co-presences for negotiation market data.

\vspace{6pt}

\begin{figure*}[ht]
\begin{center}
\includegraphics[width=1\linewidth]{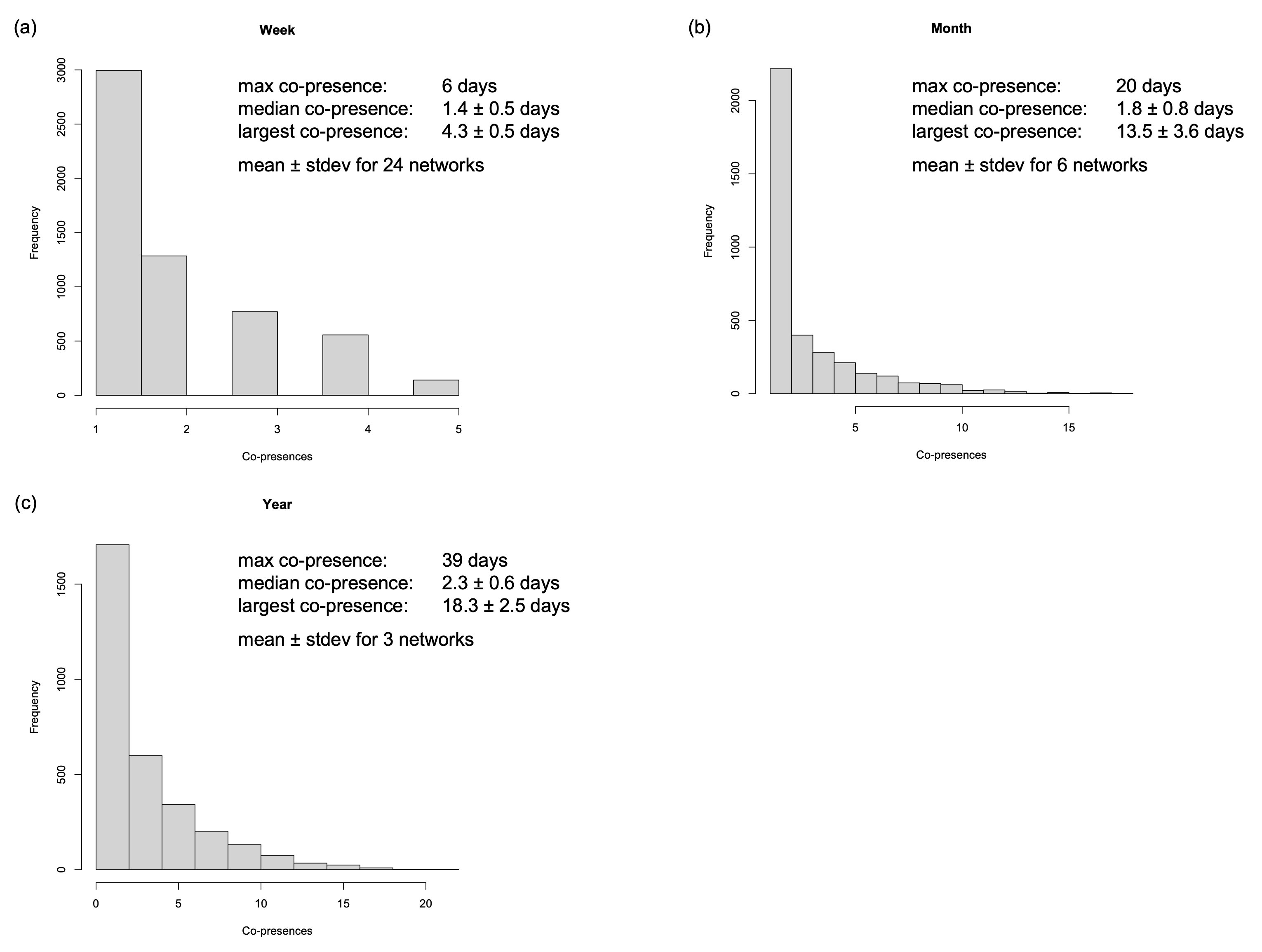}
\end{center}
\caption{Histograms of the relative frequency of co-presences for plant-pollinator data~\cite{CaraDonnaData}. Panels show co-presences for all networks combined at the specific level of temporal aggregation while statistics for the median and largest number of co-presences statistics present averages and standard deviations for individual networks.
\label{FigureS4}}
\end{figure*}

\noindent \textbf{Figure \ref{FigureS4}.} Histograms of the relative frequency of co-presences for plant-pollinator data.

\clearpage

\begin{table*}[t]
\footnotesize
\caption{Proportion of empirical nestedness generated by the degree distribution ($\Delta\tilde{\rho}_{\text{DD}}$) and phenology ($\Delta\tilde{\rho}_{\text{Phen}}$) models at three levels of temporal aggregation: weekly (W), monthly (M), and annual (Y); mean and standard deviation (in parentheses) of 1000 realizations.
\label{TableS1}}
\begin{center}
\begin{tabular}{lrrll}
Data set       & Ref                    & Agg & $\Delta\tilde{\rho}_{\text{DD}}$     & $\Delta\tilde{\rho}_{\text{Phen}}$     \\
Auction        & \cite{FishMarketData}  & Y   & 0.26 (0.02) & 0.91 (0.02) \\
               &                        & M   & 0.24 (0.03) & 0.53 (0.03) \\
               &                        & W   & 0.26 (0.05) & 0.30 (0.05) \\
Negotiation    & \cite{FishMarketData}  & Y   & 0.26 (0.03) & 0.81 (0.03) \\
               &                        & M   & 0.27 (0.03) & 0.44 (0.03) \\
               &                        & W   & 0.28 (0.04) & 0.28 (0.04) \\
CaraDonna      & \cite{CaraDonnaData}   & Y   & 0.29 (0.09) & 0.78 (0.10) \\
               &                        & M   & 0.32 (0.13) & 0.72 (0.14) \\
               &                        & W   & 0.43 (0.70) & 0.52 (0.74) \\
Alarcon2008    & \cite{Alarcon2008}     & Y   & 0.36 (0.08) & 0.63 (0.18) \\
Benadi2014     & \cite{Benadi2014}      & Y   & 0.27 (0.02) & 0.83 (0.03) \\
               &                        & M   & 0.37 (0.03) & 0.52 (0.04) \\
Burkle2012     & \cite{Burkle2012}      & Y   & 0.28 (0.05) & 0.54 (0.05) \\
Chacoff2018    & \cite{Chacoff2018}     & Y   & 0.31 (0.09) & 0.73 (0.10) \\
Fruend2010     & \cite{Fruend2010}      & Y   & 0.29 (0.05) & 1.03 (0.06) \\
               &                        & M   & 0.34 (0.13) & 0.92 (0.17) \\
LaraRomero2016 & \cite{Lara-Romero2016} & Y   & 0.30 (0.06) & 0.82 (0.07) \\
LeBuhnYY       & \cite{PlantPollData}   & Y   & 0.29 (0.07) & 0.89 (0.08) \\
Olito2015      & \cite{Olito2015}       & Y   & 0.33 (0.08) & 1.02 (0.09) \\
Rasmussen2013  & \cite{Rasmussen2013}   & Y   & 0.31 (0.09) & 1.01 (0.11) \\
ResascoYY      & \cite{PlantPollData}   & Y   & 0.37 (0.10) & 0.71 (0.11) \\
Simanonok2014  & \cite{Simanonok2014}   & Y   & 0.29 (0.06) & 0.93 (0.07) \\
Thompson2018   & \cite{Thompson2018}    & Y   & 0.35 (0.07) & 0.99 (0.08) \\
Vazquez2003    & \cite{Vazquez2003}     & Y   & 0.44 (0.13) & 0.92 (0.12) \\
               &                        & M   & 0.67 (1.38) & 1.95 (3.71) \\
Weiner2014     & \cite{Weiner2014}      & Y   & 0.29 (0.02) & 0.81 (0.02) \\
               &                        & M   & 0.35 (0.05) & 0.62 (0.05) \\
               &                        & W   & 0.39 (0.10) & 0.42 (0.09) \\
Winfree2014    & \cite{Winfree2014}     & Y   & 0.31 (0.10) & 0.67 (0.11) \\
WinfreeYYa     & \cite{WinfreeYYa}      & Y   & 0.30 (0.08) & 0.82 (0.09) \\
               &                        & M   & 0.48 (0.79) & 0.86 (1.21) \\
WinfreeYYb     & \cite{WinfreeYYa}      & Y   & 0.29 (0.06) & 0.83 (0.07) \\
               &                        & M   & 0.43 (0.54) & 0.76 (0.61) \\
WinfreeYYc     & \cite{PlantPollData}   & Y   & 0.35 (0.14) & 0.50 (0.14) \\
WinfreeYYd     & \cite{WinfreeYYd}      & Y   & 0.26 (0.03) & 0.80 (0.03) \\
               &                        & M   & 0.27 (0.05) & 0.63 (0.05) \\
               &                        & W   & 0.31 (0.11) & 0.40 (0.11) \\
WinfreeYYe     & \cite{PlantPollData}   & Y   & 0.30 (0.07) & 0.96 (0.10) \\
               &                        & M   & 0.30 (0.08) & 0.96 (0.10) \\
               &                        & W   & 0.41 (0.39) & 0.75 (0.43) \\
WinfreeYYf     & \cite{WinfreeYYf}      & Y   & 0.27 (0.06) & 0.54 (0.07) \\
               &                        & M   & 0.38 (0.29) & 0.51 (0.29)
\end{tabular}
\end{center}
\end{table*}

\clearpage
\begin{table*}[t]
\footnotesize
\caption{Phi coefficients ($\phi$) and F-scores ($F_1$) for the degree distribution (DD) and phenology (Phen) models at three levels of temporal aggregation: weekly (W), monthly (M), and annual (Y); mean and standard deviation (in parentheses) of 1,000 realizations.
\label{TableS2}}
\begin{center}
\begin{tabular}{lrrllll}
Data set       & Ref                  & Agg & $\phi_{\text{DD}}$     & $\phi_{\text{Phen}}$ & ${F_1}_{\text{DD}}$ & ${F_1}_{\text{Phen}}$    \\
Auction        & \cite{FishMarketData}  & Y   & 0.25 (0.01) & 0.47 (0.01) & 0.53 (0.01) & 0.67 (0.00) \\
               &                        & M   & 0.20 (0.01) & 0.29 (0.01) & 0.41 (0.01) & 0.47 (0.01) \\
               &                        & W   & 0.17 (0.02) & 0.14 (0.02) & 0.35 (0.01) & 0.33 (0.01) \\
Negotiation    & \cite{FishMarketData}  & Y   & 0.23 (0.01) & 0.35 (0.01) & 0.61 (0.00) & 0.68 (0.00) \\
               &                        & M   & 0.17 (0.01) & 0.16 (0.01) & 0.38 (0.01) & 0.37 (0.01) \\
               &                        & W   & 0.13 (0.01) & 0.08 (0.01) & 0.26 (0.01) & 0.22 (0.01) \\
CaraDonna      & \cite{CaraDonnaData}   & Y   & 0.14 (0.03) & 0.34 (0.03) & 0.26 (0.02) & 0.44 (0.02) \\
               &                        & M   & 0.17 (0.04) & 0.28 (0.04) & 0.31 (0.03) & 0.41 (0.03) \\
               &                        & W   & 0.20 (0.08) & 0.16 (0.07) & 0.38 (0.06) & 0.35 (0.06) \\
Alarcon2008    & \cite{Alarcon2008}     & Y   & 0.10 (0.02) & 0.17 (0.02) & 0.16 (0.02) & 0.22 (0.02) \\
Benadi2014     & \cite{Benadi2014}      & Y   & 0.07 (0.01) & 0.19 (0.01) & 0.10 (0.01) & 0.22 (0.01) \\
               &                        & M   & 0.08 (0.02) & 0.14 (0.02) & 0.14 (0.02) & 0.19 (0.02) \\
Burkle2012     & \cite{Burkle2012}      & Y   & 0.12 (0.02) & 0.19 (0.02) & 0.19 (0.02) & 0.25 (0.02) \\
Chacoff2018    & \cite{Chacoff2018}     & Y   & 0.11 (0.02) & 0.19 (0.03) & 0.19 (0.02) & 0.27 (0.02) \\
Fruend2010     & \cite{Fruend2010}      & Y   & 0.09 (0.01) & 0.28 (0.02) & 0.14 (0.01) & 0.32 (0.02) \\
               &                        & M   & 0.08 (0.05) & 0.21 (0.05) & 0.16 (0.04) & 0.28 (0.05) \\
LaraRomero2016 & \cite{Lara-Romero2016} & Y   & 0.15 (0.02) & 0.31 (0.02) & 0.29 (0.02) & 0.42 (0.02) \\
LeBuhnYY       & \cite{PlantPollData}   & Y   & 0.07 (0.02) & 0.29 (0.02) & 0.12 (0.02) & 0.33 (0.02) \\
Olito2015      & \cite{Olito2015}       & Y   & 0.09 (0.02) & 0.26 (0.02) & 0.14 (0.02) & 0.30 (0.02) \\
Rasmussen2013  & \cite{Rasmussen2013}   & Y   & 0.10 (0.02) & 0.28 (0.02) & 0.19 (0.02) & 0.35 (0.02) \\
ResascoYY      & \cite{PlantPollData}   & Y   & 0.10 (0.02) & 0.20 (0.03) & 0.17 (0.02) & 0.26 (0.02) \\
Simanonok2014  & \cite{Simanonok2014}   & Y   & 0.10 (0.02) & 0.26 (0.02) & 0.16 (0.02) & 0.31 (0.02) \\
Thompson2018   & \cite{Thompson2018}    & Y   & 0.11 (0.02) & 0.26 (0.02) & 0.19 (0.02) & 0.33 (0.02) \\
Vazquez2003    & \cite{Vazquez2003}     & Y   & 0.13 (0.03) & 0.49 (0.03) & 0.24 (0.03) & 0.55 (0.02) \\
               &                        & M   & 0.24 (0.10) & 0.39 (0.09) & 0.45 (0.06) & 0.55 (0.06) \\
Weiner2014     & \cite{Weiner2014}      & Y   & 0.11 (0.01) & 0.28 (0.01) & 0.15 (0.01) & 0.31 (0.01) \\
               &                        & M   & 0.12 (0.02) & 0.20 (0.01) & 0.17 (0.01) & 0.25 (0.02) \\
               &                        & W   & 0.13 (0.03) & 0.14 (0.03) & 0.21 (0.02) & 0.21 (0.02) \\
Winfree2014    & \cite{Winfree2014}     & Y   & 0.09 (0.02) & 0.17 (0.03) & 0.17 (0.02) & 0.24 (0.02) \\
WinfreeYYa     & \cite{WinfreeYYa}      & Y   & 0.11 (0.02) & 0.28 (0.03) & 0.18 (0.02) & 0.34 (0.02) \\
               &                        & M   & 0.12 (0.05) & 0.21 (0.05) & 0.23 (0.04) & 0.31 (0.04) \\
WinfreeYYb     & \cite{WinfreeYYa}      & Y   & 0.09 (0.02) & 0.25 (0.02) & 0.15 (0.02) & 0.30 (0.02) \\
               &                        & M   & 0.10 (0.05) & 0.21 (0.05) & 0.21 (0.04) & 0.30 (0.04) \\
WinfreeYYc     & \cite{PlantPollData}   & Y   & 0.23 (0.05) & 0.24 (0.05) & 0.39 (0.04) & 0.40 (0.04) \\
WinfreeYYd     & \cite{WinfreeYYd}      & Y   & 0.16 (0.01) & 0.35 (0.01) & 0.26 (0.01) & 0.42 (0.01) \\
               &                        & M   & 0.16 (0.02) & 0.25 (0.02) & 0.26 (0.02) & 0.35 (0.02) \\
               &                        & W   & 0.14 (0.03) & 0.15 (0.03) & 0.25 (0.03) & 0.26 (0.03) \\
WinfreeYYe     & \cite{PlantPollData}   & Y   & 0.12 (0.03) & 0.27 (0.03) & 0.23 (0.02) & 0.36 (0.02) \\
               &                        & M   & 0.12 (0.03) & 0.28 (0.03) & 0.23 (0.02) & 0.36 (0.02) \\
               &                        & W   & 0.11 (0.04) & 0.15 (0.05) & 0.22 (0.04) & 0.26 (0.04) \\
WinfreeYYf     & \cite{WinfreeYYf}      & Y   & 0.08 (0.02) & 0.23 (0.02) & 0.15 (0.02) & 0.29 (0.02) \\
               &                        & M   & 0.11 (0.05) & 0.19 (0.05) & 0.23 (0.04) & 0.29 (0.04)
\end{tabular}
\end{center}
\end{table*}

\clearpage
\begin{table*}[t]
\footnotesize
\caption{Backbone of nestedness results for the phenology model at an annual level of temporal aggregation: (i) ratio of nestedness values for the subsets of true positive and false negative links, $\rho_{\text{TP}} / \rho_{\text{FN}}$ and (ii) ratio of the deviations of nestedness relative to comparable Erd\"os-R\'enyi random graphs, $\rho_{\text{TP},\Delta\text{ER}} / \rho_{\text{FN,ER}}$, where $\rho_{\text{TP},\Delta\text{ER}} = (\rho_{\text{TP}} - \rho_{\text{TP,ER}})/(1 - \rho_{\text{TP,ER}})$ is the proportional deviation for the subset of true positive links and $\rho_{\text{FN,ER}} = (\rho_{\text{FN}} - \rho_{\text{FN,ER}})/(1 - \rho_{\text{FN, RG}})$ is the proportional deviation for the subset of false negative links; mean and standard deviation (in parentheses) of 1000 realizations.
\label{TableS3}}
\begin{center}
\begin{tabular}{lrll}
Data set       & Ref                    & $\rho_{\text{TP}} / \rho_{\text{FN}}$ & $\rho_{\text{TP},\Delta\text{RG}} / \rho_{\text{FN},\Delta\text{RG}}$ \\
Auction        & \cite{FishMarketData}  & 1.64 (0.02)   & 2.60 (0.17) \\
Negotiation    & \cite{FishMarketData}  & 1.55 (0.01)   & 2.56 (0.23) \\
CaraDonna      & \cite{CaraDonnaData}   & 1.29 (0.06)   & 2.02 (0.64) \\
Alarcon2008    & \cite{Alarcon2008}     & 1.26 (0.09)   & 0.92 (0.48) \\
Benadi2014     & \cite{Benadi2014}      & 1.27 (0.04)   & 1.24 (0.12) \\
Burkle2012     & \cite{Burkle2012}      & 1.16 (0.06)   & 0.89 (0.23) \\
Chacoff2018    & \cite{Chacoff2018}     & 1.17 (0.07)   & 0.77 (0.48) \\
Fruend2010     & \cite{Fruend2010}      & 1.38 (0.06)   & 1.56 (0.25) \\
LaraRomero2016 & \cite{Lara-Romero2016} & 1.31 (0.04)   & 2.25 (0.50) \\
LeBuhnYY       & \cite{PlantPollData}   & 1.46 (0.08)   & 2.04 (0.54) \\
Olito2015      & \cite{Olito2015}       & 1.46 (0.07)   & 1.98 (0.49) \\
Rasmussen2013  & \cite{Rasmussen2013}   & 1.37 (0.07)   & 2.25 (0.83) \\
ResascoYY      & \cite{PlantPollData}   & 1.22 (0.08)   & 0.86 (0.56) \\
Simanonok2014  & \cite{Simanonok2014}   & 1.38 (0.07)   & 1.77 (0.40) \\
Thompson2018   & \cite{Thompson2018}    & 1.22 (0.06)   & 1.55 (0.38) \\
Vazquez2003    & \cite{Vazquez2003}     & 1.07 (0.06)   & 1.60 (0.76) \\
Weiner2014     & \cite{Weiner2014}      & 1.29 (0.03)   & 1.31 (0.08) \\
Winfree2014    & \cite{Winfree2014}     & 1.32 (0.10)   & 1.58 (0.60) \\
WinfreeYYa     & \cite{WinfreeYYa}      & 1.28 (0.07)   & 1.46 (0.44) \\
WinfreeYYb     & \cite{WinfreeYYa}      & 1.18 (0.06)   & 0.96 (0.27) \\
WinfreeYYc     & \cite{PlantPollData}   & 1.13 (0.07)   & 0.97 (0.85) \\
WinfreeYYd     & \cite{WinfreeYYd}      & 1.41 (0.04)   & 1.77 (0.17) \\
WinfreeYYe     & \cite{PlantPollData}   & 1.33 (0.06)   & 1.73 (0.51) \\
WinfreeYYf     & \cite{WinfreeYYf}      & 1.09 (0.06)   & 0.75 (0.28)
\end{tabular}
\end{center}
\end{table*}

\clearpage
\begin{table*}[t]
\footnotesize
\caption{Proportion of empirical nestedness ($\Delta\tilde{\rho}$), phi coefficients ($\phi$), and F-scores ($F_1$) for the degree distribution (DD), presence (Pres), and phenology (Phen) models at an annual level of temporal aggregation, mean and standard deviation (in parentheses) of 1000 realizations.
\label{TableS4}}
\begin{center}
\begin{adjustbox}{angle=90}
\begin{tabular}{lrlllllllll}
Data set       & Ref                    & $\Delta\tilde{\rho}_{\text{DD}}$ & $\Delta\tilde{\rho}_{\text{Pres}}$ & $\Delta\tilde{\rho}_{\text{Phen}}$ & $\phi_{\text{DD}}$ & $\phi_{\text{Pres}}$ & $\phi_{\text{Phen}}$ & ${F_1}_{\text{DD}}$ & ${F_1}_{\text{Pres}}$ & ${F_1}_{\text{Phen}}$ \\
Auction        & \cite{FishMarketData}  & 0.26 (0.02)                      & 1.04 (0.02)                        & 0.91 (0.02)                        & 0.25 (0.01)        & 0.31 (0.01)          & 0.47 (0.01)          & 0.53 (0.01)         & 0.58 (0.00)           & 0.67 (0.00)           \\
Negotiation    & \cite{FishMarketData}  & 0.26 (0.03)                      & 0.37 (0.03)                        & 0.81 (0.03)                        & 0.23 (0.01)        & 0.22 (0.01)          & 0.35 (0.01)          & 0.61 (0.00)         & 0.61 (0.00)           & 0.68 (0.00)           \\
CaraDonna      & \cite{CaraDonnaData}   & 0.30 (0.09)                      & 0.59 (0.09)                        & 0.78 (0.10)                        & 0.14 (0.03)        & 0.25 (0.03)          & 0.34 (0.03)          & 0.26 (0.02)         & 0.36 (0.02)           & 0.44 (0.02)           \\
Alarcon2008    & \cite{Alarcon2008}     & 0.36 (0.08)                      & 0.56 (0.08)                        & 0.63 (0.08)                        & 0.10 (0.02)        & 0.14 (0.02)          & 0.17 (0.02)          & 0.16 (0.02)         & 0.20 (0.02)           & 0.22 (0.02)           \\
Benadi2014     & \cite{Benadi2014}      & 0.27 (0.02)                      & 0.59 (0.03)                        & 0.83 (0.03)                        & 0.07 (0.01)        & 0.14 (0.01)          & 0.19 (0.01)          & 0.10 (0.01)         & 0.17 (0.01)           & 0.22 (0.01)           \\
Burkle2012     & \cite{Burkle2012}      & 0.28 (0.05)                      & 0.47 (0.06)                        & 0.54 (0.05)                        & 0.12 (0.02)        & 0.15 (0.02)          & 0.19 (0.02)          & 0.19 (0.02)         & 0.22 (0.02)           & 0.25 (0.02)           \\
Chacoff2018    & \cite{Chacoff2018}     & 0.31 (0.09)                      & 0.58 (0.09)                        & 0.73 (0.10)                        & 0.11 (0.02)        & 0.14 (0.02)          & 0.19 (0.03)          & 0.19 (0.02)         & 0.22 (0.02)           & 0.27 (0.02)           \\
Fruend2010     & \cite{Fruend2010}      & 0.29 (0.05)                      & 0.84 (0.06)                        & 1.03 (0.06)                        & 0.09 (0.01)        & 0.24 (0.02)          & 0.28 (0.02)          & 0.14 (0.01)         & 0.28 (0.02)           & 0.32 (0.02)           \\
LaraRomero2016 & \cite{Lara-Romero2016} & 0.30 (0.06)                      & 0.51 (0.06)                        & 0.82 (0.07)                        & 0.15 (0.02)        & 0.23 (0.02)          & 0.31 (0.02)          & 0.29 (0.02)         & 0.35 (0.02)           & 0.42 (0.02)           \\
LeBuhnYY       & \cite{PlantPollData}   & 0.29 (0.07)                      & 1.01 (0.08)                        & 0.89 (0.08)                        & 0.07 (0.02)        & 0.25 (0.02)          & 0.29 (0.02)          & 0.12 (0.02)         & 0.29  (0.02)          & 0.33 (0.02)           \\
Olito2015      & \cite{Olito2015}       & 0.33 (0.08)                      & 0.80 (0.08)                        & 1.02 (0.09)                        & 0.09 (0.02)        & 0.20 (0.02)          & 0.26 (0.02)          & 0.14 (0.02)         & 0.25 (0.02)           & 0.30 (0.02)           \\
Rasmussen2013  & \cite{Rasmussen2013}   & 0.31 (0.09)                      & 0.83 (0.10)                        & 1.01 (0.11)                        & 0.10 (0.02)        & 0.21. (0.02)         & 0.28 (0.02)          & 0.19 (0.02)         & 0.29 (0.02)           & 0.35 (0.02)           \\
ResascoYY      & \cite{PlantPollData}   & 0.37 (0.10)                      & 0.66 (0.11)                        & 0.71 (0.11)                        & 0.10 (0.02)        & 0.17 (0.03)          & 0.20 (0.03)          & 0.17 (0.02)         & 0.23 (0.02)           & 0.26 (0.02)           \\
Simanonok2014  & \cite{Simanonok2014}   & 0.29 (0.06)                      & 0.72 (0.07)                        & 0.93 (0.07)                        & 0.10 (0.02)        & 0.21 (0.02)          & 0.26 (0.02)          & 0.16 (0.02)         & 0.27 (0.02)           & 0.31 (0.02)           \\
Thompson2018   & \cite{Thompson2018}    & 0.35 (0.07)                      & 1.02 (0.08)                        & 0.99 (0.08)                        & 0.11 (0.02)        & 0.24 (0.02)          & 0.26 (0.02)          & 0.19 (0.02)         & 0.31 (0.02)           & 0.33 (0.02)           \\
Vazquez2003    & \cite{Vazquez2003}     & 0.44 (0.13)                      & 0.85 (0.11)                        & 0.92 (0.12)                        & 0.13 (0.03)        & 0.46 (0.03)          & 0.49 (0.03)          & 0.24 (0.03)         & 0.53 (0.03)           & 0.55 (0.02)           \\
Weiner2014     & \cite{Weiner2014}      & 0.29 (0.02)                      & 0.58 (0.02)                        & 0.81 (0.02)                        & 0.11 (0.01)        & 0.21 (0.01)          & 0.28 (0.01)          & 0.15 (0.01)         & 0.24 (0.01)           & 0.31 (0.01)           \\
Winfree2014    & \cite{Winfree2014}     & 0.31 (0.10)                      & 0.67 (0.11)                        & 0.67 (0.11)                        & 0.09 (0.02)        & 0.16 (0.03)          & 0.17 (0.03)          & 0.17 (0.02)         & 0.24 (0.02)           & 0.24 (0.02)           \\
WinfreeYYa     & \cite{WinfreeYYa}      & 0.30 (0.08)                      & 0.87 (0.09)                        & 0.82 (0.09)                        & 0.11 (0.02)        & 0.26 (0.03)          & 0.28 (0.03)          & 0.18 (0.02)         & 0.32 (0.02)           & 0.34 (0.02)           \\
WinfreeYYb     & \cite{WinfreeYYa}      & 0.29 (0.06)                      & 0.89 (0.07)                        & 0.83 (0.07)                        & 0.09 (0.02)        & 0.22 (0.02)          & 0.25 (0.02)          & 0.15 (0.02)         & 0.28 (0.02)           & 0.30 (0.02)           \\
WinfreeYYc     & \cite{PlantPollData}   & 0.35 (0.14)                      & 0.33 (0.13)                        & 0.50 (0.14)                        & 0.23 (0.05)        & 0.16 (0.05)          & 0.24 (0.05)          & 0.39 (0.04)         & 0.33 (0.04)           & 0.40 (0.04)           \\
WinfreeYYd     & \cite{WinfreeYYd}      & 0.26 (0.03)                      & 0.50 (0.03)                        & 0.80 (0.03)                        & 0.16 (0.01)        & 0.22 (0.01)          & 0.35 (0.01)          & 0.26 (0.01)         & 0.31 (0.01)           & 0.42 (0.01)           \\
WinfreeYYe     & \cite{PlantPollData}   & 0.30 (0.07)                      & 0.64 (0.09)                        & 0.96 (0.10)                        & 0.12 (0.03)        & 0.19 (0.03)          & 0.27 (0.03)          & 0.23 (0.02)         & 0.29 (0.02)           & 0.36 (0.02)           \\
WinfreeYYf     & \cite{WinfreeYYf}      & 0.27 (0.06)                      & 0.84 (0.09)                        & 0.54 (0.07)                        & 0.08 (0.02)        & 0.23 (0.02)          & 0.23 (0.02)          & 0.15 (0.02)         & 0.29 (0.02)           & 0.29 (0.02)          
\end{tabular}
\end{adjustbox}
\end{center}
\end{table*}
\clearpage

\newpage

\newpage

\newpage
\bibliographystyle{Style}

\let\oldthebibliography\thebibliography
\let\endoldthebibliography\endthebibliography
\renewenvironment{thebibliography}[1]{
  \begin{oldthebibliography}{#1}
    \setlength{\itemsep}{0em}
    \setlength{\parskip}{0em}
}
{
  \end{oldthebibliography}
}

\bibliography{literature}